\documentclass[aip,apl,reprint]{revtex4-1} 

\usepackage[T1]{fontenc}
\usepackage{graphicx}
\usepackage{units}
\usepackage{amsmath,amssymb}

\begin{document}

\title{Dry-transferred CVD graphene for inverted spin valve devices}

\author{Marc Dr\"ogeler}
\affiliation{JARA-FIT and 2nd Institute of Physics, RWTH Aachen University, 52074 Aachen, Germany}
\author{Luca Banszerus}
\affiliation{JARA-FIT and 2nd Institute of Physics, RWTH Aachen University, 52074 Aachen, Germany}
\author{Frank Volmer}
\affiliation{JARA-FIT and 2nd Institute of Physics, RWTH Aachen University, 52074 Aachen, Germany}
\author{Takashi Taniguchi}
\affiliation{National Institute for Materials Science, 1-1 Namiki, Tsukuba 305-0044, Japan}
\author{Kenji Watanabe}
\affiliation{National Institute for Materials Science, 1-1 Namiki, Tsukuba 305-0044, Japan}
\author{Bernd Beschoten}
\affiliation{JARA-FIT and 2nd Institute of Physics, RWTH Aachen University, 52074 Aachen, Germany}
\author{Christoph Stampfer}
\email[]{stampfer@physik.rwth-aachen.de}
\affiliation{JARA-FIT and 2nd Institute of Physics, RWTH Aachen University, 52074 Aachen, Germany}
\affiliation{Peter Gr\"unberg Institute (PGI-9), Forschungszentrum J\"ulich, 52425 J\"ulich, Germany}

\date{\today}

\begin{abstract}
Integrating high-mobility graphene grown by chemical vapor deposition (CVD) into spin transport devices is one of the key tasks in graphene spintronics. We use a van der Waals pickup technique to transfer CVD graphene by hexagonal boron nitride (hBN) from the copper growth substrate onto predefined Co/MgO electrodes to build inverted spin valve devices.
Two approaches are presented: (i) a process where the CVD-graphene/hBN stack is first patterned into a bar and then transferred by a second larger hBN crystal onto spin valve electrodes and (ii) a direct transfer of a CVD-graphene/hBN stack. We report record high spin lifetimes in CVD graphene of up to $\unit[1.75]{ns}$ at room temperature. Overall, the performances of our devices are comparable to devices fabricated from exfoliated graphene also revealing nanosecond spin lifetimes. We expect that our dry transfer methods pave the way towards more advanced device geometries not only for spintronic applications but also for CVD-graphene-based nanoelectronic devices in general where patterning of the CVD graphene is required prior to the assembly of  final van der Waals heterostructures.

\end{abstract}


\maketitle
Dry transfer techniques have become a key technology for building high-quality van der Waals heterostructures from 2D materials.\cite{Wang01112013,doi:10.1021/nn400280c,doi:10.1021/nl5006542,Geim2013,2053-1583-1-1-011002,PRL.99.232104} For example, this technique allows to fully encapsulate graphene between two insulating hexagonal boron nitride (hBN) crystals serving as excellent substrates for graphene.\cite{NatNano.5.722} Additionally, the graphene gets fully protected against air contaminations and is not exposed to wet chemistry which allows to unveil unique electronic properties with charge carrier mobilities exceeding $\unit[1,000,000]{cm^2/(Vs)}$ at cryogenic temperatures.\cite{Wang01112013} This technique has also been adopted for single and bilayer graphene grown by chemical vapor deposition (CVD)\cite{Banszeruse1500222,schmitz2017high} and it was demonstrated that the CVD graphene transport properties are equivalent to devices based on high-mobility exfoliated graphene.\cite{Banszeruse1500222,doi:10.1021/acs.nanolett.5b04840,schmitz2017high}

In graphene spintronics,\cite{GrapheneSpintronics,2DMaterials.Roche} there has also been a tremendous improvement in spin and charge transport properties when using the van der Waals pick-up approach to protect graphene by hBN.\cite{Droegeler2014,doi:10.1021/acs.nanolett.6b00497,NanoLett.16.4825,Avsar2016,PhysRevLett.113.086602,PhysRevB.86.161416,PhysRevB.93.115441} The best device performance so far has been achieved for \textit{exfoliated} graphene which gets picked up by a large hBN crystal and is subsequently transferred on prepatterned Co/MgO electrodes yielding spin lifetimes of $\tau_\text{s}=\unit[12]{ns}$, spin diffusion lengths of $\lambda_\text{s}=\unit[30]{\mu m}$ combined with charge carrier mobilities of more than $\mu = \unit[20,000]{cm^2/(Vs)}$ at room temperature.\cite{doi:10.1021/acs.nanolett.6b00497}

In contrast, state-of-the-art CVD graphene-based spin valve devices exhibit minor charge and spin transport properties with spin lifetimes typically below $\unit[1]{ns}$ and carrier mobilities up to $\unit[2,000]{cm^2/(Vs)}$.\cite{Avsar2011,Kamalakar2015,PhysRevApplied.6.054015} This is mainly caused by the wet transfer process which is used to delaminate CVD graphene from the growth substrate (Cu foil). Longer spin lifetimes ($\tau_\text{s}=\unit[1.2]{ns}$) could only be measured in devices with long spin transport channel lengths ($L = \unit[16]{\mu m}$) where the influence of contact-induced spin scattering \cite{PhysRevB.86.235408,2053-1583-2-2-024001,Amamou2016,PhysRevApplied.6.054015} gets diminished.

In this Letter we show inverted spin valve devices based on dry-transferred CVD graphene. First (i) we utilize a two-step fabrication process providing bar-shaped CVD graphene: Therefore we use a first hBN crystal to delaminate CVD graphene (Gr) from the Cu foil (Fig.~1a). As the CVD graphene gets ripped along the outer edges of the hBN crystal, this CVD-graphene/hBN stack is transferred onto a Si/SiO$_2$ substrate where we etch a well-defined bar (Fig.~1b) through the whole heterostructure by reactive ion etching (RIE). In a next step we pick up the CVD-graphene/hBN bar with a second hBN crystal (Fig.~1c) and place the whole stack onto prepatterned Co/MgO electrodes (see Figs.~1d,1f). Throughout the whole process the CVD graphene is protected by the hBN against solvents, polymers and the metal hard mask which is used for etching. We also use a second approach (ii) where the picked-up CVD-graphene/hBN stack is directly transferred onto the spin valve electrodes (see Fig.~1g). All devices show comparable spin transport properties as devices based on exfoliated graphene.

\begin{figure*}
\includegraphics[width=0.95\linewidth]{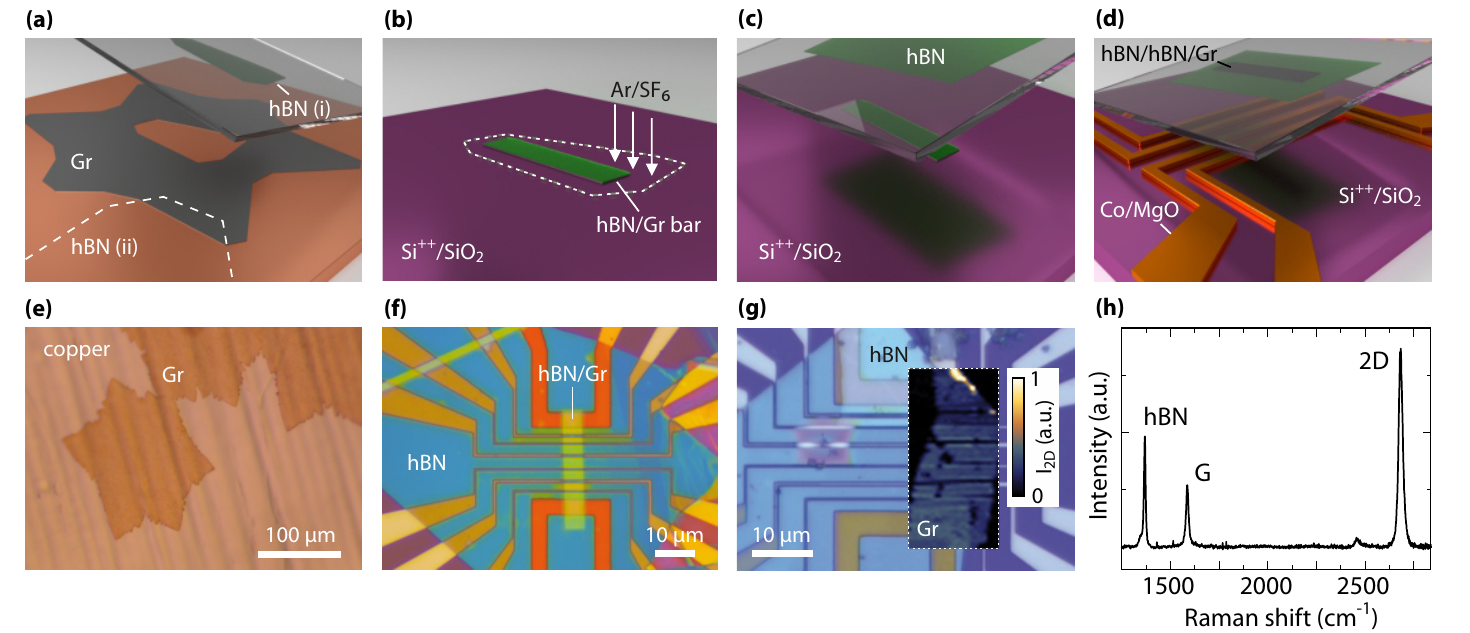}%
\caption{(a-d) Illustration of the fabrication process. (a) After the growth of the CVD graphene on a copper foil it is picked up with hBN supported by PDMS/PVA/PMMA. (b) The CVD-graphene/hBN stack is deposited on Si/SiO$_2$, the polymers are dissolved and the stack (dotted line) is etched by RIE to pattern the desired shape. (c) The etched hBN/graphene bar is picked up using a second hBN crystal.
(d) The polymer/hBN/hBN/CVD-graphene stack is placed on the Co/MgO electrodes. Thereafter, the polymer membrane is dissolved.
(e) Optical image of CVD graphene on the copper foil after oxidation of the graphene-to-copper interface.
(f) Optical image of a finished CVD graphene spin valve fabricated by the etching technique. (g) Optical image of a CVD graphene spin valve device fabricated by the stamping technique where the graphene is ripped from the CVD flake along the outer edges of a hBN stamp (see dashed line in panel (a)) and directly deposited on the prepatterned Co/MgO electrode structure.
A Raman 2D-peak intensity map (see white dashed box) is added to depict the position and shape of the CVD graphene after transfer. (h) Averaged Raman spectrum of the device shown in panel (g).}
\label{fig:process}
\end{figure*}

We fabricate CVD-graphene-based inverted spin valves by first defining magnetic electrodes on a Si$^{++}$/SiO$_2$ ($\unit[285]{nm}$) substrate (see Fig.~1d) using standard electron beam lithography (EBL), molecular beam epitaxy (MBE) and a lift-off process. They consist of $\unit[35]{nm}$ Co followed by $\unit[1.2]{nm}$ of MgO. The latter is used as a spin injection and detection barrier. Graphene is grown by low pressure chemical vapor deposition on the inside of copper enclosures which typically results in large graphene flakes of up to a few hundred micrometers (see Fig.~1e). Process details are given in Ref.~[\citenum{Banszeruse1500222}]. The flakes are stored under ambient conditions for some days to allow for oxidation along the copper-to-graphene interface. This oxidation weakens the graphene-to-copper adhesion which allows subsequent pick-up of the CVD graphene by hBN. \cite{Banszeruse1500222} We first focus on the process to fabricate bar-shape CVD-graphene/hBN structures. For this, we pick-up the CVD graphene from the copper foil (Fig.~1a) by using hBN which was exfoliated onto a stack of polydimethylsiloxane (PDMS), polyvinylalcohol (PVA) and polymethylmethacrylate (PMMA). Thereafter, we place the graphene/hBN stack on a clean Si/SiO$_2$ chip and dissolve the polymer stack in deionized (DI) water, acetone and isopropanol. The desired shape of the graphene/hBN stack is defined by an aluminum hardmask made by EBL and a lift-off process and subsequently etched by RIE in an Ar/SF$_6$ plasma (Fig.~1b). After the etching, the aluminum is removed using tetramethylammonium hydroxide (TMAH) and thereafter washed with DI water. During the whole process of patterning the graphene is protected by the overlying hBN crystal. For the final transfer we exfoliate an additional hBN flake on a PMMA membrane which we use to pick up the graphene/hBN bar (Fig.~1c). In a last step the graphene/hBN/hBN stack is deposited onto the Co/MgO electrodes (Fig.~1d) and the polymer membrane is dissolved in acetone and isopropanol. An optical image of a finished device is shown in Fig.~1f. The yellow stripe in the center of the electrode structure represents the patterned graphene/hBN bar which was transferred by a larger hBN crystal which shows up as the blue large flake.

We note that the pick-up of the etched graphene/hBN stack is not as simple as for unpatterned graphene/hBN, which likely may be caused by residues stemming either from the hard mask or from organic solvents which are present on top of the hBN. Apparently, this leads to an overall weaker adhesion to the hBN which is used for the pick-up. Another observation is that the supporting PMMA membrane starts to corrugate when it gets retracted from the substrate during the pick-up of the graphene/hBN stack (Fig.~1c). We suspect that the membrane has a strong adhesion to the substrate due the etching of its surface by the RIE process (Fig.~1b) which results in irreversible stretching when retracting the membrane. The corrugations in the membrane may prevent that the hBN/hBN/graphene smoothly adapt to the electrode structure of the spin valve devices (Fig.~1d). Hence, the hBN/hBN/graphene stacks may float away or get partially lifted when the membrane is dissolved.  During this procedure residues from the fabrication process of the electrodes may bond to graphene and reduce its electronic quality.To allow for a better adhesion of the supporting hBN crystal to both the electrodes and the substrates we heat the whole sample beyond the glass temperature of the PMMA ($T\approx\unit[105]{^\circ C}$) prior to the dissolving of the membrane.
\begin{figure}
\includegraphics[width=0.95\linewidth]{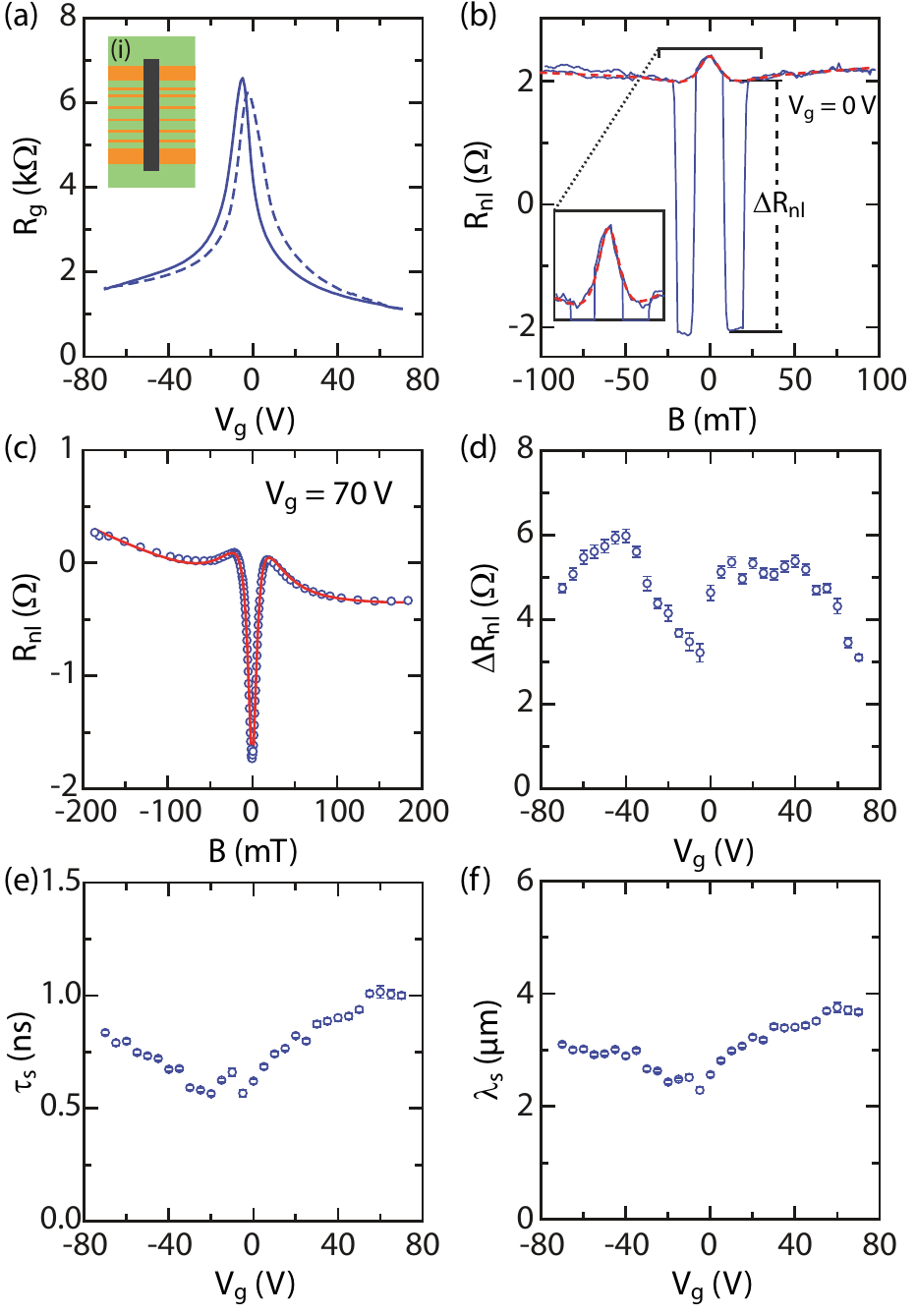}%
\caption{(a) Graphene resistance vs. gate voltage for a CVD graphene device fabricated by the etching technique (i) where the trace (solid line) is measured from $V_\text{g}=-70 \text{ to } \unit[70]{V}$ and the retrace (dashed line) from $V_\text{g}=70 \text{ to } \unit[-70]{V}$. The inset shows a schematic of the device with the etched hBN/CVD graphene bar (black), the supporting large hBN crystal (green) and the Co/MgO electrodes (orange) (b) Room temperature spin valve measurement. The red dashed line shows a fit by the standard Hanle model for spin precession. The inset shows a close-up of the peak at low B-fields. (c) Hanle curves for anti-parallel alignment at $V_\text{g} = \unit[70]{V}$ with Hanle fit (red curve) including a second-order background function. (d-f) Spin amplitude, spin lifetime and spin diffusion length vs. gate voltage. All data were taken at room temperature.}
\label{fig:results_etching}
\end{figure}

We next focus of typical charge and spin transport properties obtained from such devices. All measurements were carried out under vacuum conditions at room temperature using standard low frequency lock-in techniques.\cite{PhysRevB.88.161405} The measured gate dependent graphene resistance is shown in Fig.~\ref{fig:results_etching}a. From this curve we extract a charge carrier mobility of $\mu=\unit[3,000]{cm^2/(Vs)}$. In contrast, devices where we did not need to apply the additional heating procedure exhibit mobilities of $\mu=\unit[10,000-20,000]{cm^2/(Vs)}$. However, the latter set of devices suffer from an improper mechanical contact of the hBN/hBN/graphene stack to the substrate which results in maximum spin lifetimes of only several hundred picoseconds most likely due to the ingress of solvents during the final fabrication step. This general trend was also observed in devices with exfoliated graphene.~\cite{doi:10.1021/acs.nanolett.6b00497}

Fig.~\ref{fig:results_etching}b shows spin valve measurements recorded on a spin transport region with $L=\unit[5]{\mu m}$. The data were taken in non-local geometry with an in-plane magnetic field applied along the electrode direction. The spin amplitude $\Delta R_\text{nl}$, which is given by the non-local resistance between parallel and anti-parallel alignments of the magnetization direction of the spin injection and detection electrodes, typically is on the order of several Ohms. Next to the switching there is a small peak around zero magnetic field which is symmetric in magnetic field. We attribute this to Hanle spin precession
caused by an electrode magnetization direction which is not collinear with the electrode direction. The red curve in Fig.~\ref{fig:results_etching}b is a fit by the standard Hanle model~\cite{doi:10.1021/acs.nanolett.6b00497,PhysRevB.37.5312,ISI000249789600001} and describes the data quite well even towards larger fields. It is now interesting to compare the extracted spin lifetime of $\tau_\text{s}=\unit[651]{ps}$ and the spin diffusion length of $\lambda_\text{s}=\unit[2.2]{\mu m}$ with values obtained from standard Hanle spin precession measurements, where the magnetic field is applied perpendicular to the graphene sheet. Fig.~\ref{fig:results_etching}c shows a typical Hanle curve for anti-parallel alignment of the magnetization directions. Gate dependent spin amplitude $\Delta R_\text{nl}$, spin lifetime $\tau_\text{s}$ and spin diffusion length $\lambda_\text{s}$ are depicted in Figs.~\ref{fig:results_etching}d-f. The spin amplitude first increases away from the CNP and decreases again for high values of the gate voltage $V_\text{g}$. As for most of the exfoliated samples both $\tau_\text{s}$ and $\lambda_\text{s}$ increase for larger gate voltages,\cite{doi:10.1021/acs.nanolett.6b00497,PhysRevLett.113.086602,PhysRevB.88.161405,PhysRevLett.107.047207,doi:10.1021/nl301567n,PhysRevLett.107.047206} i.e. larger carrier densities with the former reaching a maximum value of $\tau_\text{s}=\unit[1.0]{ns}$. We note that the respective spin transport parameters for $V_\text{g}=\unit[0]{V}$ match quite well to the above values which confirms our notion that the peak in the spin valve measurement indeed can be interpreted as a Hanle curve caused by perpendicularly oriented magnetic domains. They may have been formed when the sample was heated during the deposition of the hBN/hBN/graphene stack. The magnetization configuration is rather stable and cannot be changed even in large magnetic fields of up to $\unit[1]{T}$. This important finding indicates that the in-plane and out-of-plane spin lifetimes are equal and excludes significant influences of Rashba spin orbit coupling (SOC) in this device. A similar conclusion was obtained from spin transport studies on exfoliated graphene flakes where the spin lifetime anisotropy was determined by spin precession in a tilted magnetic field.\cite{Raes2016} Our results demonstrate that the transfer and etching technique allows to achieve nanosecond spin lifetimes even for CVD graphene devices with short spin transport lengths.

\begin{figure}
\centering
\includegraphics[width=0.95\linewidth]{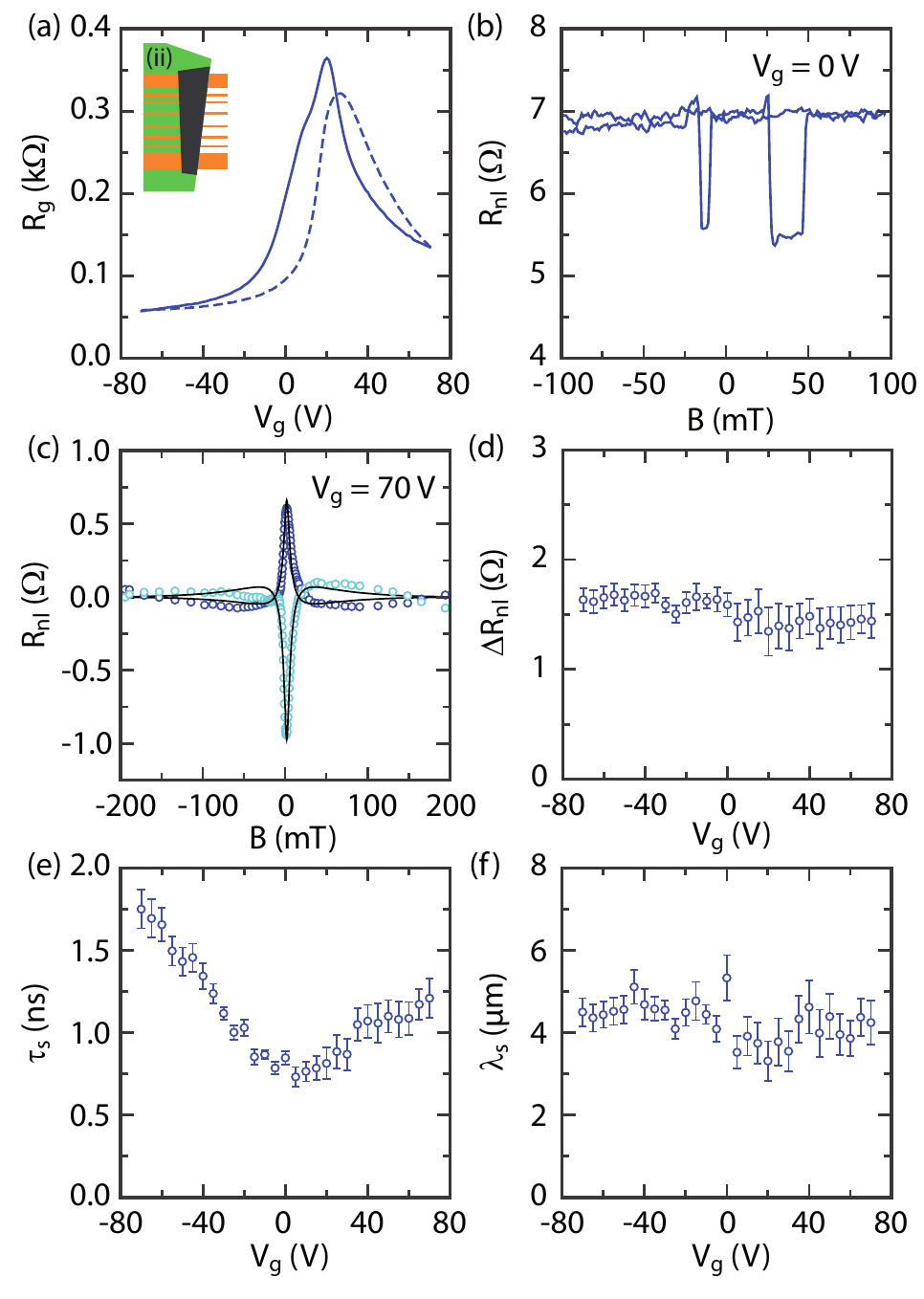}%
\caption{(a) Graphene resistance vs. gate voltage for a CVD graphene device fabricated by the stamping technique where the trace (solid line) is measured from $V_\text{g}=-70 \text{ to } \unit[70]{V}$ and the retrace (dashed line) from $V_\text{g}=70 \text{ to } \unit[-70]{V}$. The inset shows a schematic of the device where the CVD graphene got stamped by an hBN crystal (green) which are both placed onto predefined Co/MgO electrodes. (b) Room temperature spin valve measurement. (c) Hanle curves for parallel and anti-parallel alignment at $V_\text{g} = \unit[70]{V}$. The solid lines are fits by a standard Hanle spin precession model. (d-f) Spin amplitude, spin lifetime and spin diffusion length vs. gate voltage. All data were taken at room temperature.}
\label{fig:results_stamping}
\end{figure}

We also investigated a second (ii) fabrication method which we call the stamping technique. Here, we mechanically stamp-out a CVD graphene stripe from the edge of the flake. Again, we first exfoliate hBN on a stack of PMMA/PVA/PDMS. For the stamping the respective edges of the CVD and the hBN crystal are aligned in such a way that the overlap region of both flakes defines a stripe in the CVD graphene (see dashed line in Fig.~1a and label "hBN (ii)"). The pick-up of the stripe is feasible since there is only a strong adhesion between graphene and hBN but not between PMMA and graphene. Thereafter, the stack of PDMS/PVA/PMMA/hBN/graphene is aligned and transferred to a predefined Co/MgO electrode structure. In a last step the whole stack is put in acetone and isopropanol to remove the polymer.

An optical image of the final device is shown in Fig.~1g. Additionally, we include an intensity map of the Raman 2D-peak of the CVD graphene to identify both the position and the shape of the graphene. The averaged spectrum of the Raman map is depicted in Fig.~1h exhibiting a 2D line width of $\unit[20.5]{cm}^{-1}$ demonstrating the high electronic quality of the transferred CVD graphene.\cite{2053-1583-4-2-025030}

In Fig.~\ref{fig:results_stamping}a we plot the graphene resistance as a function of back gate voltage for the device shown in Fig.~\ref{fig:process}g. The room temperature electron mobility $\mu = \unit[7,000]{cm^2/(Vs)}$ is larger than for the device in Fig.~\ref{fig:results_etching} and is in the same order as for inverted spin valve devices with exfoliated graphene.
The spin valve curve and typical Hanle spin precession curves for both parallel and antiparallel alignments are shown in  Fig.~\ref{fig:results_stamping}b and Fig.~\ref{fig:results_stamping}c, respectively. We note that there is a small contribution from an antisymmetric Hanle curve which is expected if the magnetization directions of both spin injection and detection electrodes are not perfectly colinear. We consider this contribution when fitting the curves by the standard Hanle model.\cite{doi:10.1021/acs.nanolett.6b00497} The gate dependence of the extracted $\Delta R_\text{nl}$, $\tau_\text{s}$ and $\lambda_\text{s}$ values are shown in Figs.~\ref{fig:results_stamping}d-f. Both the spin amplitude and the spin diffusion length are almost independent of the gate voltage while the spin lifetime again shows the well-know V-shape dependence with spin lifetimes reaching $\tau_\text{s}=\unit[1.75]{ns}$. This value is the longest spin lifetime measured for CVD graphene by now and compares well to the average values obtained for inverted spin valve devices where we used exfoliated flakes.\cite{doi:10.1021/acs.nanolett.6b00497}

In summary, we have demonstrated two dry transfer methods for integrating CVD graphene in inverted spin valve devices. Both allow for spin lifetimes in the nanosecond regime at room temperature and high charge carrier mobilities. Although both methods still have room for improvements our findings show that the device performance of CVD-graphene-based spin transport devices
is well comparable with that of state-of-the-art exfoliated graphene-based devices. Most importantly, the nanosecond spin lifetimes indicate that there is no significant spin scattering at, for examples, copper residues on the surface of the transferred CVD graphene which might be expected to result from CVD growth in the copper enclosures.\cite{Balakrishnan2014A,PhysRevB.95.035402} We expect that the presented fabrication methods pave the way towards more advanced device geometries not only for spintronic applications but also for CVD-graphene-based nanoelectronic devices in general.


\begin{acknowledgments}
We acknowledge funding from the European Union Seventh Framework Programme under Grant Agreement No. 604391 Graphene Flagship and the Deutsche Forschungsgemeinschaft (BE 2441/9-1) and support by the Helmholtz Nano Facility (HNF)\cite{HNF} at the Forschungszentrum J\"ulich. Growth of hexagonal  boron nitride crystals was supported by the Elemental Strategy Initiative conducted by the MEXT, Japan and JSPS KAKENHI Grant Numbers JP26248061, JP15K21722 and JP25106006.

\end{acknowledgments}

%

\end{document}